\documentclass[%
aip,
rsi,
amsmath,amssymb,
reprint%
,longbibliography
]{revtex4-2}
\usepackage{graphicx}

\usepackage{textcomp}
\usepackage{units}
\usepackage[breaklinks,pdfusetitle,pdfauthor={Paul van der Hulst}]{hyperref}	

\begin{document}
	
\title{Frequency Shift Algorithm: Design of a Baseband Phase Locked Loop for Frequency-Domain Multiplexing Readout of X-ray Transition-Edge Sensor Microcalorimeters}

\author{Paul van der Hulst}
\email{p.van.der.hulst@sron.nl}
\affiliation{SRON Netherlands Institute for Space Research, Sorbonnelaan 2, 3584CA Utrecht, The Netherlands}
\author{Jan van der Kuur}
\affiliation{SRON Netherlands Institute for Space Research, Kapteynborg, Landleven 12, 9747 AD Groningen, The Netherlands}
\author{Ad Nieuwenhuizen}
\author{Davide Vaccaro}
\author{Hiroki Akamatsu}
\author{Patrick van Winden}
\author{Bert-Joost van Leeuwen}
\author{Jan-Willem den Herder}
\affiliation{SRON Netherlands Institute for Space Research, Sorbonnelaan 2, 3584CA Utrecht, The Netherlands}

\date{\today}

\begin{abstract}
	The Transition-Edge Sensor (TES) is an extremely sensitive device which is used to measure the energy of individual X-ray photons. 
	For astronomical spectrometry applications, 
	SRON develops a Frequency Domain Multiplexing (FDM) read-out system for kilopixel arrays of such TESs.
	Each TES is voltage biased at a specific frequency in the range 1 to \unit[5]{MHz}.
	Isolation between the individual pixels is obtained through very narrow-band (high-Q) lithographic LC resonators.
	To prevent energy resolution degradation due to intermodulation line noise, the bias frequencies are distributed on a regular grid.
	The requirements on the accuracy of the LC resonance frequency are very high.
	The deviation of the resonance frequencies due to production tolerances is significant with respect to the bandwidth, and a controller is necessary to compensate for the LC series impedance.
	We present two such controllers: a simple orthogonal proportional-integrating (PI) controller and a more complex impedance estimator.
	Both controllers operate in baseband and try to make the TES current in-phase with the bias voltage, effectively operating as phase-locked loops (PLL).
	They allow off-LC-resonance operation of the TES pixels, while preserving TES thermal response and energy resolution.
	Extensive experimental results -- published in a companion paper recently -- with the proposed methods, show that these controllers allow the preservation of single pixel energy resolution in multiplexed operation.
\end{abstract}

\maketitle

\section{Introduction}

Transition Edge Sensors (TES) are very sensitive radiation detectors used in ground- and space-based astronomy.\cite{dobbs_frequency_2012,adams_micro-x_2020} 
When used as bolometers, they can sense very weak signals such as far infrared radiation.\cite{jackson_spica-safari_2012}
For more energetic (e.g.\ x-ray) photons, they can be used as microcalorimeters to measure the energy of individual photons, which allows the study of the hot gas distribution in the universe, and black holes.\cite{barret_athena_2016}

TESs operate at the edge of superconductivity, an effect which manifests itself at cryogenic temperatures. 
The critical temperature for the TESs developed at SRON is typically around $\unit[100]{mK}$.
In order to limit wiring complexity for a kilopixel TES array, as well as the amount of heat produced by cold stage pre-amplifiers, and thus the required cooling power, the readout electronics multiplex the signals of multiple sensors over a single transmission channel. 
Different multiplexing strategies are available for multiplexing of TES sensors with Superconducting QUantum Interference Device (SQUID) amplifiers: Time Division Multiplexing (TDM),\cite{doriese_developments_2016} Code Division Multiplexing (CDM)\cite{stiehl_code-division_2012} and Frequency Division Multiplexing (FDM). 

Over the last few years, SRON has actively been developing FDM readout of TES sensors for the X-IFU instrument on the Athena satellite.\cite{jackson_focal_2016,den_hartog_frequency_2011}
FDM is currently the the backup readout method for X-IFU. 
The SAFARI spectrometer on the recently canceled Spica spacecraft\cite{jackson_spica-safari_2012} was also planned to be equipped with SRON FDM readout.
Recently published experimental results\cite{vaccaro_frequency_2021} show that the new techniques presented in this paper, raise the performance of FDM readout to a level competitive with the current state of the art.

The FDM method biases each TES with an ac voltage of a specific frequency. 
LC filters in series with each TES select the carrier for the TESs.
For a carrier signal at the LC resonance frequency, the impedance of the resonator is zero, biasing the TES with the provided carrier voltage.
The carrier signals for the other sensors are outside the pass-band of the LC resonator and thus they are attenuated.

Due to fabrication tolerances, the resonance frequency of the LC filters exhibits a spread of a few kHz from the intended \unit[100]{kHz} grid.\cite{bruijn_tailoring_2014}
When each pixel is operated at exactly its corresponding LC resonance frequency, the carriers are distributed somewhat randomly in frequency space.
Non-linearities in the active part of the readout chain will cause intermodulation distortion.
These distortion components can -- and will -- end up in the signal band of some pixels and interference between this frequency and the carrier will result in beating, i.e.\ modulation of the bias power.
The pixel readout signal then becomes a function of the phase of the beating, and the resulting energy resolution has been observed to deteriorate by a few electronvolt.\cite{akamatsu_progress_2020}

For many bolometer applications the problem is insignificant because the signal bandwidth is rather low.
This reduces the chances of intermodulation products to end up within this signal bandwidth.
When it does occur, a very small frequency shift relative to the LC filter bandwidth (which does not significantly change the LC series impedance) is sufficient to move it out of the signal band.
For microcalorimeters and bolometers with higher thermal bandwidth,\cite{montgomery_performance_2021} the required frequency shift to  move the distortion lines out of the pass band of the system is proportionally larger.
This increases the series impedance of the LC filter.
As a result, the TES current is no longer primarily determined by its resistance, but increasingly by the reactive series impedance of the LC circuit, which is dominant outside the resonator bandwidth.

Several methods have been proposed in the past to compensate reactive series impedance.
Digitally Enhanced Voltage Bias\cite{de_haan_improved_2012} compensates a fixed series impedance of the TES, but requires prior measurement of the series impedance, and does not use the phase information of the measured  current which is needed to separate the resistive and reactive components of the currents.
Automatic compensation of off-resonance series impedance (with an active integrator controller) has also been proposed and demonstrated.\cite{van_der_kuur_active_2018,akamatsu_progress_2020}
The primary aim of that solution was to maintain a low source impedance for the TES bias source by shifting the resonance frequency of the closed loop system.
The controllers presented in this paper are based on a reformulated control goal and have a more robust implementation in baseband.
Instead of shifting the resonance frequency, we designed an adaptive controller which provides the required reactive voltage to compensate the off-resonance LC series impedance under arbitrary frequency shift.

\section{System configuration}
\begin{figure}[htbp]
	\centering
	\includegraphics[width=\columnwidth]{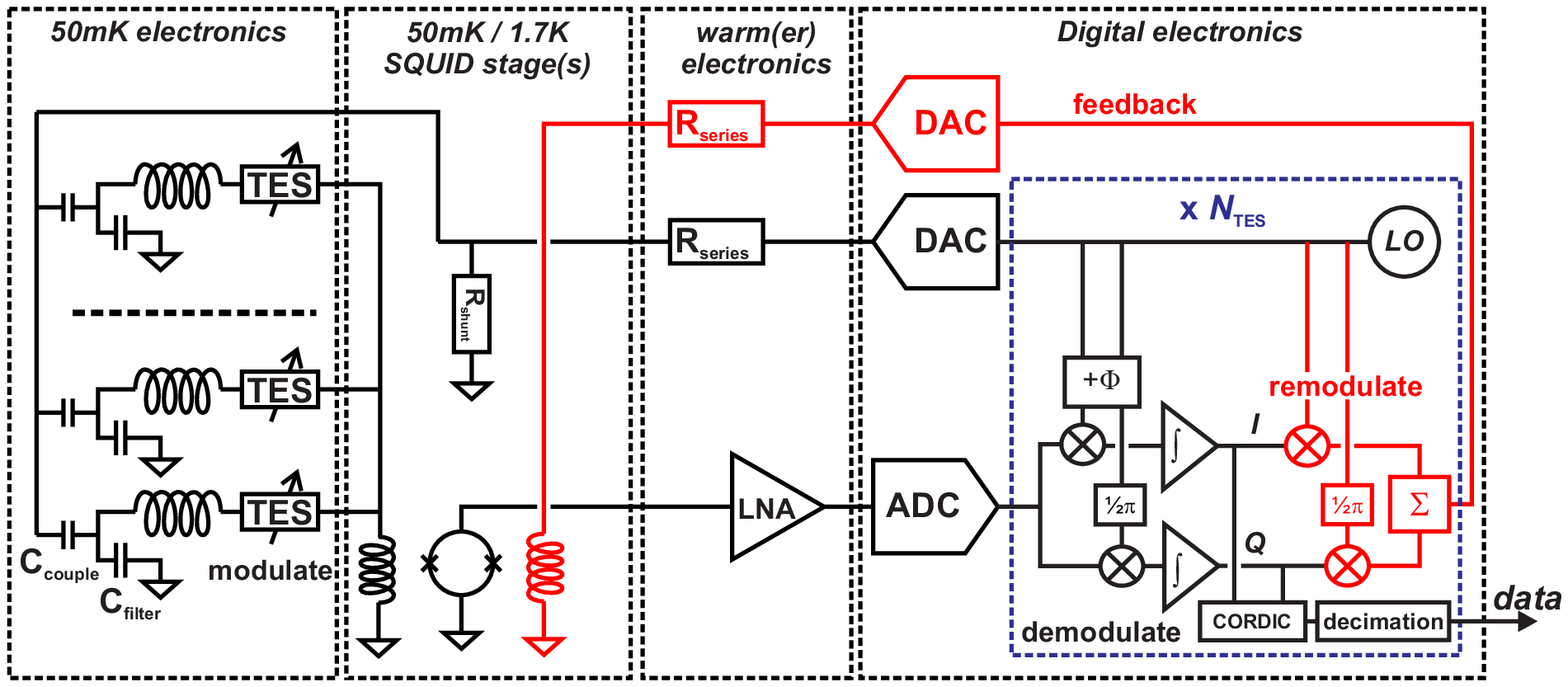}
	\caption{The FDM and BBFB system diagram for one channel, before modification. 
		The demodulation stage for data output is also used to generate the baseband feedback (in red) which linearizes the SQUID.
		Reproduced from \protect\nocite{den_hartog_low-noise_2012}
		{{den Hartog}}\ \emph {et~al.}, \href
		{https://doi.org/10.1007/s10909-012-0577-8}{{J. of Low Temp. Phys.}}, \textbf{167}, 652, 2012; licensed under a Creative Commons Attribution (CC BY) license.}
	\label{fig:den_Hartog-2012-low-noise}
\end{figure}

The configuration of the original TES readout system is shown in Fig.~\ref{fig:den_Hartog-2012-low-noise}.
The TES sensors, LC separation filters, a shunt resistor and a SQUID buffer amplifier are located at two different temperature levels in the cryostat.
The warm (analog) electronics buffers the signals to and from the cryostat, the latter through a Low Noise Amplifier (LNA).
The ac bias and feedback signals are effectively converted to current sources through series resistors.
The digital electronics contains generation of the individual ac bias signals, based on a local oscillator (LO), as well as the demodulation logic to separate the readout signals of the individual TES sensors.
The BaseBand FeedBack (BBFB) control loop\cite{takei_squid_2009,den_hartog_baseband_2009} (in red) efficiently reuses the demodulated signals to minimize the SQUID input signal level.
This feedback loop improves linearity of the SQUID buffer stage, while at the same time it acts as a band-limiting filter for the scientific output signals.

\section{Controller design}
\label{sec:Controller design}
\begin{figure}[htbp]
	\centering
	\includegraphics[scale=0.8]{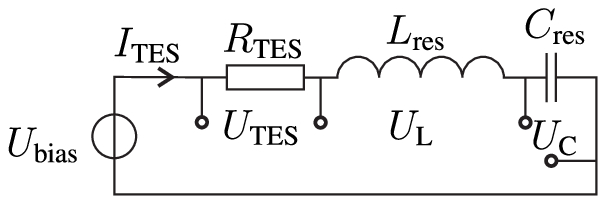}
	\caption{Fundamental single pixel TES readout circuit consisting of a bias voltage source, a TES and an LC series resonator for blocking other frequencies}
	\label{fig:LCR-circuit_landscape}
\end{figure}

\begin{figure}[htbp]
	\centering
	\includegraphics[scale=1]{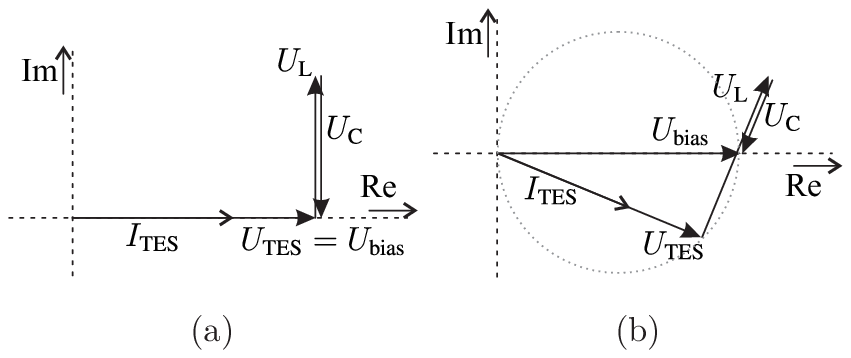}
	\caption{Phasor diagrams for the voltages in the fundamental circuit of Fig.~\ref{fig:LCR-circuit_landscape}. 
		On-resonance~(a) and off-resonance~(b)}
	\label{fig:Complex_impedance}
\end{figure}

The conceptual model of the TES circuit is shown in Fig.~\ref{fig:LCR-circuit_landscape}.
The ac voltage source $U_\text{bias}$ provides the bias voltage for the TES resistance $R_\text{TES}$.
A series LC resonant circuit in series with the TES provides frequency selectivity.
Phasor diagrams showing the voltages and current in the circuit are shown in Fig.~\ref{fig:Complex_impedance}.
When the bias frequency is equal to the LC resonance frequency, the voltages across the capacitor and coil cancel exactly as shown in Fig.~\ref{fig:Complex_impedance}a.
When the circuit is operated above resonance, the impedance of the coil increases, while the impedance of the capacitor decreases.
The voltage across the coil is thus higher than the voltage across the capacitor.
This is illustrated in the phasor diagram of Fig.~\ref{fig:Complex_impedance}b.
The capacitor and coil voltages are necessarily perpendicular to the (TES-)current.
The component voltages can be geometrically constructed with the help of the dotted circle because any triangle with the circle's diameter as one side and the 3\textsuperscript{rd}\ corner elsewhere on the circle is necessarily a right triangle. 
The bias voltage is now no longer across the TES because there is a voltage drop across the capacitor and inductor.
The TES voltage is reduced in amplitude and lags the bias voltage.

Apart from the effect of bias voltage reduction, the TES is also no longer biased with a low source impedance.
The output impedance of the bias voltage source is increased with the LC impedance.
This is problematic because the TES has a very steep positive temperature coefficient.
When a photon is captured, the TES's temperature and thus the resistance increases.
The current should now decrease, or else self-heating power of the TES increases and thermal runaway can occur.
To prevent this, the current must dominantly be determined by the TES resistance.
The output impedance of the bias source must therefore be lower than the TES resistance to make the thermal feedback loop stable.\cite{irwin_transition-edge_2005}

\subsection{Controller properties}
\subsubsection{Orthogonal output}
\begin{figure}[htbp]
	\centering
	\includegraphics[scale=1]{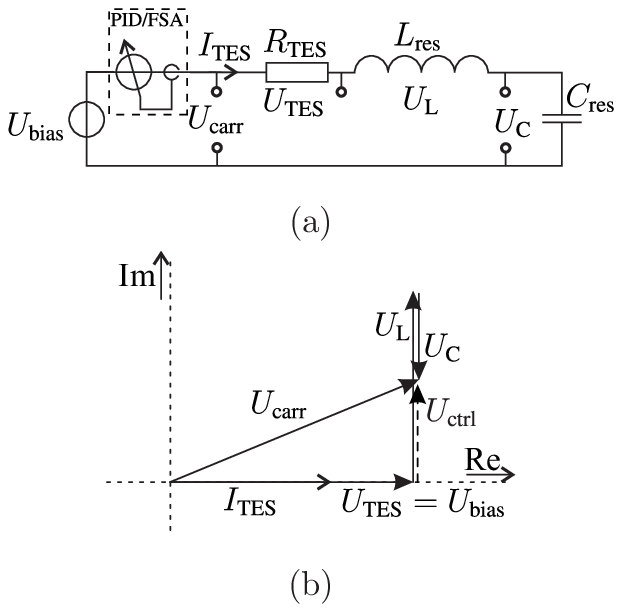}
	\caption{Fundamental circuit of the TES readout with controller~(a) and corresponding phasor diagram (b) illustrating the concept of a controller with output orthogonal to the TES current/voltage}
	\label{fig:Orthogonal control}
\end{figure}

Based on the phasor diagrams in Fig.~\ref{fig:Complex_impedance}, we can list some desired properties of the controller output voltage.
The controller should compensate the voltage drop across the LC resonator when operated off-resonance.
Since the LC impedance causes an error voltage proportional to- and perpendicular to the TES current, the compensation voltage should be
\begin{enumerate}
	\item proportional to the TES current, 
	\item \unit[90]{\textdegree} phase shifted with respect to the TES current or voltage.
\end{enumerate}
The controller is inserted into the circuit in Fig.~\ref{fig:Orthogonal control}.
The controller indicated by the dashed box in Fig.~\ref{fig:Orthogonal control}a provides the additional current-dependent voltage required to compensate the voltage across the LC resonator.
Fig.~\ref{fig:Orthogonal control}b shows the corresponding phasor diagram.
The total voltage generated by the DAC is $U_\text{carr}$, which is the sum of the orthogonal signals $U_\text{bias}$ and $U_\text{ctrl}$.
	
\subsubsection{Control target}
Fig.~\ref{fig:Complex_impedance}b and Fig.~\ref{fig:Orthogonal control}b show the difference between uncontrolled and desired off-resonance operation of the LC circuit respectively.
Only the bias voltage and TES current in these figures are known variables, either because they are output or because they are measured.

The TES current is measured with a BBFB filter\cite{takei_squid_2009} which also demodulates the signal as part of its operation.
The output of this filter is a current vector, described by in-phase ($I$) and out-of-phase $Q$(uadrature) orthogonal components. 
The $I$ and $Q$ signals, which are originally intended for scientific processing, can also be used to determine the phase shift between bias voltage and TES current.
When the TES current is in-phase with the bias voltage, the $Q$ signal is 0.
A non-zero value of the $Q$ signal corresponds to a phase-shift due to off-resonance operation. 
The $Q$ signal is therefore a convenient input for the controller, with an implicit set-point $Q_\text{ref}=0$.

\subsubsection{Controller dynamics}
When the TES current lags the bias voltage, a leading controller voltage must be provided.
The controller thus requires negative feedback.
The controlled system should also have no steady-state error. 
This implies that the controller must at least provide an integrating action.
In order to improve dynamic behavior, a proportional parallel branch is added to get a PI controller. 
Finally, to reduce interference with neighboring signals in the FDM channel due to the proportional action, the controller inputs can be filtered with two independent 1\textsuperscript{st} order low-pass filters. 

\subsection{Baseband system model}
\begin{figure}[htbp]
	\centering
	\includegraphics[scale=1]{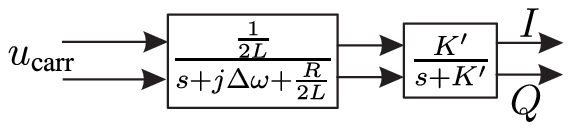}
	\caption{Complex baseband model of the LC resonator and BBFB filter}
	\label{fig:Complex_baseband_LC_BBFB}
\end{figure}

The proposed controllers will operate in baseband rather than on ac signals. 
Figure~\ref{fig:Complex_baseband_LC_BBFB} shows the baseband model of the LC resonator and BBFB filter.
The transformation of the LC filter and BBFB model to baseband and the reduction of the combination to the diagram in figure~\ref{fig:Complex_baseband_LC_BBFB} is detailed in Appendix A.

The first stage is the complex transfer function of the LC filter under frequency shift operation. 
The imaginary term $j\Delta \omega$ in the denominator indicates that the filter has a single non-conjugate complex pole, realized by a cross path between the real and imaginary signal paths of the filter.\cite{kiss_complex_2002}
An interesting feature is that since $R$ and $L$ are the same for each resonator in the 1 to \unit[5]{MHz} range, the dynamic behavior of the LC filter is not determined by the resonance frequency, just by the frequency shift.
The second stage in the diagram represents the BBFB filter, transformed for complex signals.
This transformation modifies the BBFB integrator gain $K$ to $K'=\nicefrac{K}{2}$ in the complex implementation.

The baseband model of the system models the ac amplitude and -phase response of the pixel current which carries the power information resulting from photon events.
By removing the frequency modulation from the model, the model has become much simpler.
More importantly the derivation demonstrates that the pixel behavior is independent of the carrier frequency.
In the model, the carrier- and LC resonance frequencies are exchanged for a single parameter $\Delta\omega$.
Only the frequency \emph{shift} and the bandwidths of both the LC filter and the BBFB filter determine the photon response.

An additional benefit of the baseband model over the hf model is that simulation time step can be up to \unit[1000]{\texttimes} larger, resulting in much faster simulation. 
Normal simulation of the hf model requires at least 10 intervals per period of the ac carrier ($T_\text{carr}=$ 200 to \unit[1000]{ns}) to get reasonable accuracy.
The photon events that are being studied are typically 3 orders of magnitude slower, in the order of \unit[1]{ms}, due to the limited thermal- and resonator bandwidths.

\subsection{Calibration}
To enhance clarity, the bias voltage and TES current are drawn in-phase for a resistive load in the phasor diagrams of figures~\ref{fig:Complex_impedance} and~\ref{fig:Orthogonal control}b.
This is correct when considering just the cryogenic circuit, but there is a significant additional time delay through the signal lines to and from the cryostat, as well as due to digital processing.
Due to the narrow frequency band of a pixel, this delay can be expressed as a single offset angle between the bias voltage- and current phasors.
The delay can therefore be compensated by adding the inverse angle offset between the modulation and demodulation signals.
With the delay compensated, we can pretend that the bias and measurement frames are the same.
A phase angle between voltage and current detected in the processing stage of the digital electronics is then equal to the natural phase-shift of the LC resonator.

The delay compensation angle must be calibrated individually for each carrier.
The impedance of the cryogenic circuit is resistive on resonance, and therefore bias voltage and measured current should also be in-phase.
Assuming that the bias voltage is in the $I$ direction and controller is deactivated, the residual phase angle of an imperfect delay compensation projects a fraction of the measured current on the $Q$ axis.
Calibration is than a matter of tuning the delay compensation angle until the $Q$ component in the measured current is 0.

\section{Q-nuller}

\begin{figure*}[htbp]
	\centering
	\includegraphics[scale=0.8]{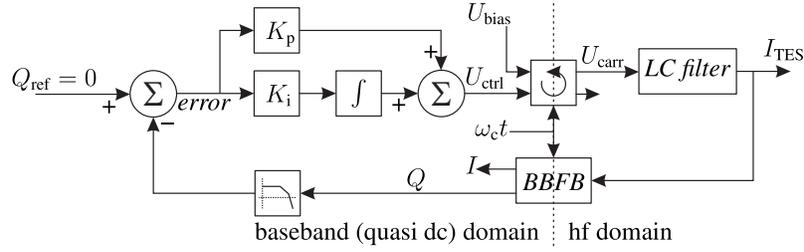}
	\caption{Q-nuller configuration of the control loop: the $Q$ output of the BBFB filter is fed to the PI controller, the controller output $U_\text{ctrl}$ is fed to the imaginary (orthogonal) input of the modulator. Modulation with the carrier frequency is indicated by a complex vector rotation, of which only the real output (the hf voltage $U_\text{carr}$) is used. Reproduced from \protect\nocite{vaccaro_frequency_2021} {Vaccaro}\ \emph {et~al.}, \href{https://doi.org/10.1063/5.0032011}{Rev. Sci. Instrum.}, \textbf{92}, {033103} {2021}, with the permission of AIP Publishing.}
	\label{fig:Q-nuller-Davide}
\end{figure*}

As a proof of concept, initially the simplest possible version of the controller was implemented.
This variant -- dubbed the `Q-nuller' -- is shown in Fig.~\ref{fig:Q-nuller-Davide}.
It simply sends the output of the PI controller directly to the carrier modulator.
In the modulator, the bias amplitude setpoint is multiplied with a cosine function because it is at the real input and the controller output is multiplied with a sine function because it is at the imaginary input.
This forces a \unit[90]{\textdegree} phase shift between the bias and controller components.
This orthogonal modulation functionality was already present in the existing modulator implementation and thus required little modification of the digital electronics firmware.

When operating off-resonance, the voltage across the LC resonator is proportional to the magnitude of the TES current.
The output of the Q-nuller controller only depends on the Q-component of this current.
The controller can react only when a change in current results in an associated angle error and thus a $Q$ current.
This makes it react slow to changes in current magnitude.
To be able to follow dynamic events (photon events), the controller gain must be high, which may affect stability.

\subsection{Stability}
Initially, only the TES readout system (LC filter and BBFB filter) with an integrating controller will be evaluated. 
Further filtering can be added if required, e.g.\ for stability.

The Q-nuller controller only operates on the imaginary components of the TES current and carrier voltage. 
It is therefore not possible to analyze the system using complex math.
The transfer of the model in Fig.~\ref{fig:Complex_baseband_LC_BBFB} must be expanded to separate real and imaginary signal paths.
\begin{figure*}[htbp]
	\centering
	\includegraphics[scale=0.8]{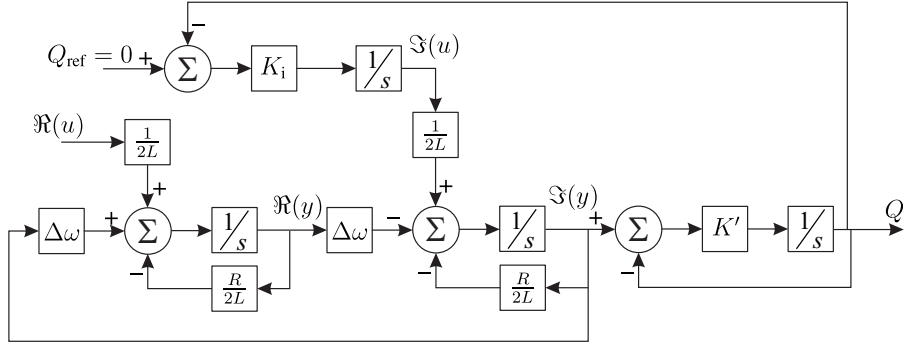}
	\caption{Baseband model of closed loop system with Q-nuller integrating controller, where $\Re(u)$ is the bias set point and $\Im(u)$ the controller value. The outputs $\Re(y)$ and $\Im(y)$ are the I- and Q-components of the TES current respectively.}
	\label{fig:FSA_Q-nuller}
\end{figure*}
The readout system controlled with the Q-nuller is shown in Fig.~\ref{fig:FSA_Q-nuller}.
The feedback control loop is only active on the imaginary signal path. 
The real signal path has no controller and/or feedback.

For stability analysis of the closed loop, the open loop baseband transfer of the system with Q-nuller controller needs to be evaluated:
\begin{align}
	H_\text{ol,bb,Q}(s)&=H_{\text{LC,}\Im}(s)\times H_\text{BBFB}(s)\times H_\text{ctrl}(s)\\
	&=\frac{\frac{1}{2L}(s+\frac{R}{2L})}{s^2 +\frac{R}{L}s + (\frac{R}{2L})^2 + \Delta\omega^2} \times \frac{K'}{s+K'}\times \frac{K_\text{i}}{s}\label{eq:Q-nuller_open_loop}
\end{align}
where $H_{\text{LC,}\Im}(s)$ is the transfer of the LC filter in baseband from $\Im(u)$ to $\Im(y)$ and $H_\text{BBFB}(s)$ and $H_\text{ctrl}(s)$ are the baseband transfer of the BBFB filter and the integrating Q-nuller controller respectively.

\begin{figure}[htbp]
	\centering
	\includegraphics[width=\columnwidth]{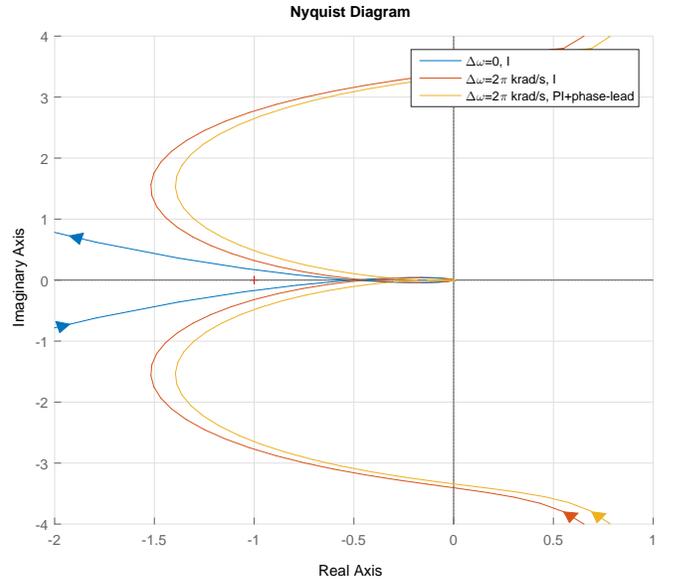}
	\caption{Nyquist plots for a system with Q-nuller controller ($H_\text{ol,bb,Q}(s)$ of \eqref{eq:Q-nuller_open_loop}, $L=\unit[2]{\text{µH}}$, $R_\text{TES}=\unit[15]{m\Omega}$, $\omega_\text{BBFB}=\unitfrac[2\pi\times 10]{krad}{s}$ and controller gain $K_\text{i}=500$).
		Curves are provided for an I-controller with $\Delta\omega=\unitfrac[0]{rad}{s}$ (blue) and $\Delta\omega=\unitfrac[2\pi\times 1]{krad}{s}$ (orange), and a PI controller with $\omega_\text{PI}=\omega_\text{BBFB}$ and $\omega_\text{LP}=3\times\omega_\text{BBFB}$(yellow).
		Gain margin decreases with frequency shift, but phase margin is improved by a phase-lead filter.}
	\label{fig:Nyquist_Q-nuller}
\end{figure}

Stability can be evaluated through the Nyquist plot of $H_\text{ol,bb,Q}(s)$ which is shown in Fig.~\ref{fig:Nyquist_Q-nuller} for 3 different configurations.
When the controller is an integrating controller as in \eqref{eq:Q-nuller_open_loop}, the blue line represents on-resonance operation -- or a frequency shift of \unit[0]{kHz} -- and the orange line a frequency shift of \unit[1]{kHz}.
In both situations this system is conditionally stable:
for high frequencies, the Nyquist plot approaches the origin from above, crossing the negative real axis between \textminus 1 and the origin.
It will become unstable when the gain is increased more than about~2.5\texttimes.
In that case the Nyquist plot will cross the negative real axis on the left from \textminus 1 and the closed loop system will be unstable.

The system can be made unconditionally stable by adding a proportional signal path to the integrating controller. 
This decreases the high frequency phase-shift by \unit[90]{\textdegree}, which causes the Nyquist plot to approach the origin from the left. 
Thus the Nyquist plot no longer \emph{crosses} the negative real axis, regardless of gain.
The location of the PI zero is typically set at a few kHz, between the LC filter bandwidth and the maximum expected frequency shift.
Unfortunately, the proportional controller action increases the magnitude of out-of-band signals.
The high frequency signal content interferes with neighboring FDM channels.
An additional low-pass filter with a crossover frequency higher than the PI zero is therefore necessary. 
The pole of this filter is placed above the unity gain frequency so its' phase shift will not affect stability.
The PI zero in combination with the low-pass filter effectively constitutes a phase-lead network added to the integrating controller.
The Nyquist plot of this situation is plotted in Fig.~\ref{fig:Nyquist_Q-nuller} in yellow.
The phase lead characteristic increases stability margins or allows the controller gain to be increased. 
That said, the system remains only \emph{conditionally} stable.

\section{Impedance estimator}
\begin{figure*}[htbp]
	\centering
	\includegraphics[scale=0.8]{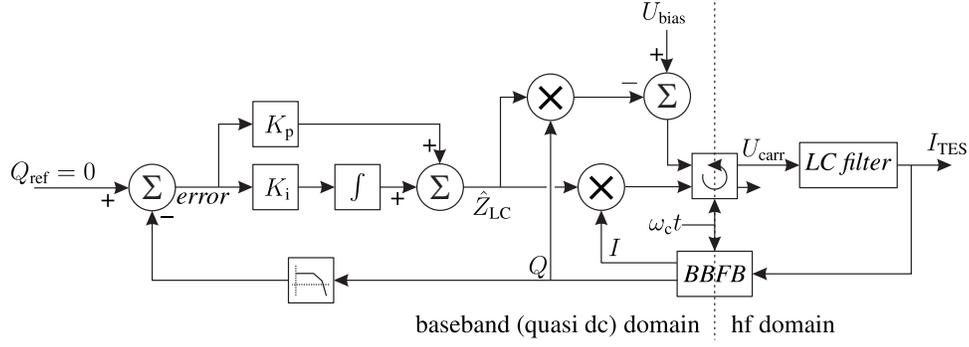}
	\caption{Z-estimator configuration of the control loop: the (PI) controller output $U_\text{ctrl}$ is multiplied with the $I$/$Q$ current components before being fed to the modulator. Modulation with the carrier frequency is indicated by a vector rotation, of which only the real output (the hf voltage $U_\text{carr}$) is used.}
	\label{fig:Z-estimator-controller-paper}
\end{figure*}
A more elaborate implementation of the controller uses the output of the PI controller as estimate of the LC impedance.
This variant is called the `Z-estimator' and is shown in Fig.~\ref{fig:Z-estimator-controller-paper}.
This estimated impedance (controller output) is multiplied with the measured current and added to the bias voltage with \unit[90]{\textdegree} phase shift.
Thus the controller voltage is a prediction of the voltage across the LC resonator which needs to be compensated, given the current.
The response of this controller is inherently fast due to the multiplication with the measured current.
The PI estimator itself can be very slow since it aims to estimate the value of a constant property.

\subsection{Stability}

\begin{figure}[htbp]
	\centering
	\includegraphics[scale=0.78]{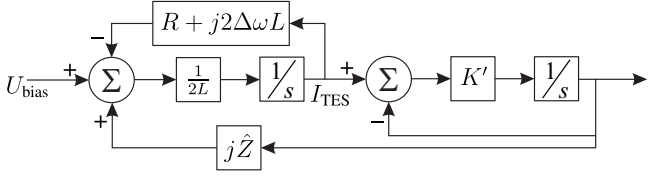}
	\caption{Baseband model for the closed loop system with Z-estimator. All signals, including $U_\text{bias}$ and $I_\text{TES}$ are complex.}
	\label{fig:Impedance_estimator_closed_loop_complex}
\end{figure}

The Z-estimator controller is a non-linear controller because it multiplies two signal paths.
However, the estimator output should be virtually constant, since it provides an estimate of a constant value: the off-resonance impedance of the LC resonator.
The controller gain can therefore be set very low, or even zero after a short initialization interval.
The open-loop baseband transfer of the system with Z-estimator controller is then:
\begin{align}
	H_\text{ol,bb,Z}(s)&=H_\text{LC}(s)\times H_\text{BBFB}(s)\times j\hat{Z}\\
	&=\frac{\frac{1}{2L}}{s +j\Delta\omega+\frac{R}{2L} } \times \frac{K'}{s+K'}\times j\hat{Z}\label{eq:Z-estimator_open_loop}\ .
\end{align}
Here, $H_\text{LC}(s)$, $H_\text{BBFB}(s)$ and $j\hat{Z}$ are the transfers for analytic baseband signals of the LC filter, the baseband feedback filter and the estimated impedance respectively.
The equivalent closed-loop model for complex signals is shown in Fig.~\ref{fig:Impedance_estimator_closed_loop_complex}.
From Fig.~\ref{fig:Impedance_estimator_closed_loop_complex} it is clear that for sufficient bandwidth of the BBFB filter, the imaginary component in the feedback will be canceled when $\hat{Z}=2\Delta\omega L$.
When this is the case, the controller successfully restores the system response to the on-resonance behavior.

\begin{figure}[htbp]
	\centering
	\includegraphics[width=\columnwidth]{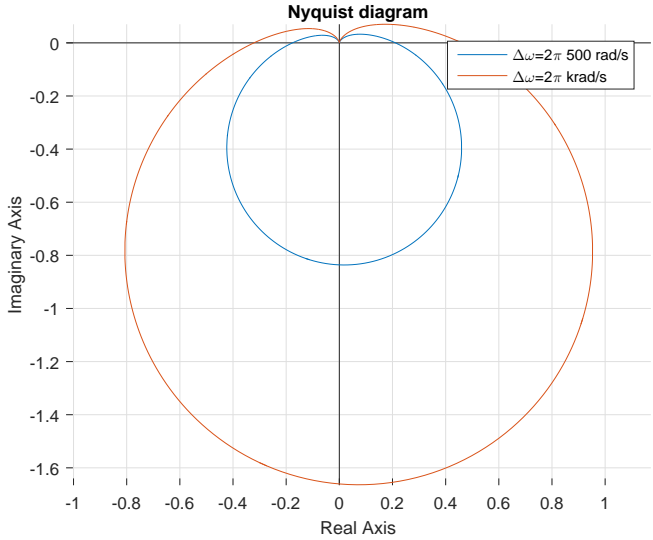}
	\caption{Nyquist diagram of the system with Z-estimator controller ($H_\text{ol,bb,Z}(s)$ of \eqref{eq:Z-estimator_open_loop}, $L=\unit[2]{\text{µH}}$, $R_\text{TES}=\unit[15]{m\Omega}$ and estimator value $\hat{Z}=2\Delta\omega L$). Curves are provided for $\Delta\omega=\unit[2\pi\times 500]{rad/s}$ (blue) and $\Delta\omega=\unit[2\pi\times 1]{krad/s}$ (orange). Although the phase margin decreases for rising $\Delta\omega$, the system is unconditionally stable for any frequency shift, provided the loop gain $\hat{Z}\le 2\Delta\omega L$}
	\label{fig:Nyquist_Z-estimator}
\end{figure}

The Nyquist plot of $H_\text{ol,bb,Z}(s)$ in shown in Fig.~\ref{fig:Nyquist_Z-estimator}.
It shows that this system is unconditionally stable.
The radius and asymmetry of the figure depend on the BBFB filter bandwidth $K'$ and the frequency shift $\Delta\omega$.
For increasing frequency shift, the -1 point is approached asymptotically.
Even for a ridiculously large value of e.g.\ $\Delta\omega=\unitfrac[2\pi\times 50]{krad}{s}$ (5\texttimes\ the BBFB filter bandwidth) the -1 point is still not encircled.

As with the Q-nuller controller, additional filtering of the measured current (the BBFB filter output) is beneficial to reduce interference with neighboring pixels.
The pole must be placed above the unity gain frequency of the loop to prevent the filter from causing instability.
A crossover frequency of \unit[25]{kHz} ($\nicefrac{1}{4}$ of the frequency separation between pixels in the FDM scheme) is considered to be a safe compromise between stability and interference.

\section{Parallel resonator leakage}

\begin{figure}[tbp]
	\centering
	\includegraphics[width=\columnwidth]{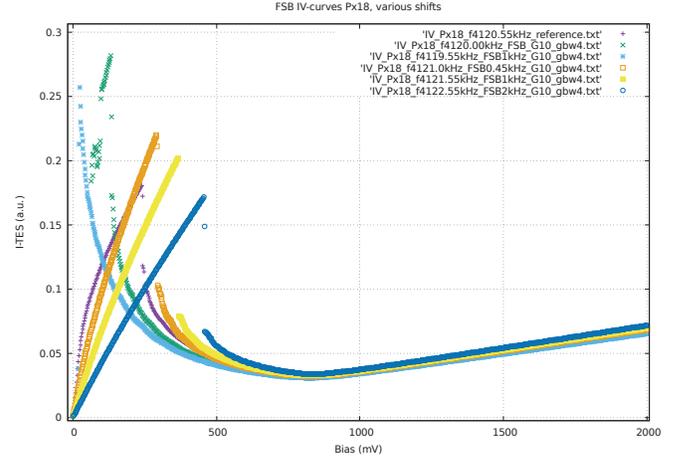}
	\caption{TES I/V curves for various frequency shifts with Q-nuller controller. The curves diverge due to current leakage through the parallel resonators.}
	\label{fig:IV curves with frequency shift}
\end{figure}

In the X-IFU FDM system, up to 40 pixels are operated in parallel, with a carrier frequency spacing of about \unit[100]{kHz}.
When operated on-resonance, the impedance at the carrier frequency is dominated by the low impedance of the TES.
The impedance of the neighbors is negligibly high.
When a pixel is operated off-resonance, the impedance of the LC resonator will increase, depending on the frequency shift.
At some point the total impedance will be noticeably influenced by the reactive impedance of the neighbors which causes an offset in the measured $Q$-current.
This results in a calibration error of the controller which attempts to zero the $Q$ current.
This error results in bad reproducibility when plotting the I/V curves, as illustrated in Fig.~\ref{fig:IV curves with frequency shift}.

\subsection{Carrier dependence}
\begin{figure}[htbp]
	\centering
	\includegraphics[width=\columnwidth]{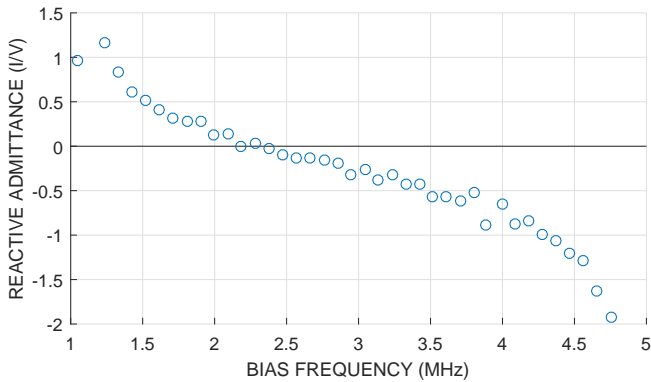}
	\caption{The total reactive admittance in parallel to each pixel as function of the pixel resonance frequency (assuming $L=\unit[2]{\text{µH}}$). The data was calculated from a list of measured resonance frequencies of the LC filter array.}
	\label{fig:Ypar_compensation_dw0_FSApaper}
\end{figure}

The effect of the neighboring resonators is a function of the carrier frequency, or rather the resonance frequencies of the neighbors.
When a neighbor pixel is tuned at a frequency above the carrier frequency, it presents a capacitive impedance because it is operated below resonance.
When a neighbor pixel is tuned at a frequency below the carrier frequency, it presents an inductive impedance because it is operated above resonance.
A low frequency pixel thus has mainly capacitive impedances in parallel whereas a high frequency pixel has mainly inductive impedances in parallel.
Necessarily the effects compensate for pixels somewhere halfway.
The calculated (imaginary) \emph{admittance} for each pixel during one of the experiments is shown in Fig.~\ref{fig:Ypar_compensation_dw0_FSApaper} as function of the pixel's LC resonance frequency.

The offset in the $Q$ current resulting from this effect can be compensated by subtracting a bias voltage dependent fraction from the measured $Q$-current. 
The compensation and its impact on the controllers are detailed in the supplemental material in appendix B.

\section{Results}
Experiments with the new controllers have been performed to test their effect on energy resolution under frequency shift, and the behavior while multiplexing multiple pixels.
For most of the experiments, the Q-nuller with PI controller was applied rather than the superior Z-estimator, for the practical reasons that it was developed earlier and that it was sufficient for the purpose of demonstrating the readout capability of FDM.
The application and optimization of the Z estimator scheme is planned for the near future.

The PI crossover zero was typically set at 2 to 3 times the pixel bandwidth as determined by the LC filter- and thermal bandwidths.
The low pass filters were configured for a bandwidth of a 2 to \unit[3]{\texttimes} $f_\text{PI}$.
These values were the same for all pixels within the 1 to \unit[5]{MHz} range of FDM carriers, which confirms our expectation from the baseband LC model that the dynamics of the pixels do not depend on the modulation frequency.

\begin{figure}[htbp]
	\centering
	\includegraphics[width=\columnwidth]{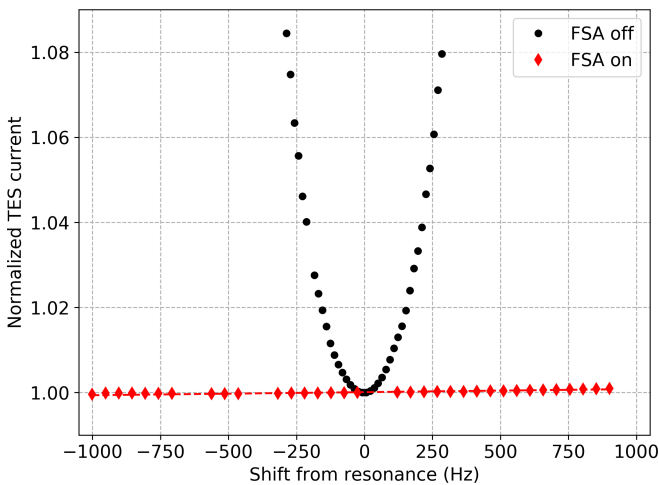}
	\caption{TES current on a typical bias point with FSA (=Q-nuller) (in)active across $\unit[\pm1]{kHz}$ frequency shift. Reproduced from \protect\nocite{vaccaro_frequency_2021} {Vaccaro}\ \emph {et~al.}, \href{https://doi.org/10.1063/5.0032011}{Rev. Sci. Instrum.}, \textbf{92}, {033103} {2021}, with the permission of AIP Publishing.}
	\label{fig:Vaccaro_2020_Application_fig9}
\end{figure}

Fig.~\ref{fig:Vaccaro_2020_Application_fig9} shows the effectiveness of the controller in providing a constant bias voltage to the TES, irrespective of the frequency shift.
When biased in the transition region, the TES behaves as an almost constant power sink as witnessed by the approximately $\nicefrac{1}{x}$ shape of the typical TES i/v curve.\cite{gottardi_ac_2011}
Because the current increases when the voltage decreases, it presents a dynamically negative resistance which is initially larger than the positive impedance of the series LC.
Therefore, the TES current increases when a frequency shift is applied and no controller is active (black circles).
Due to the negative dynamic resistance of the TES, the carrier voltage must be increased in order to decrease the current.
With the Q-nuller (the red diamonds), the carrier voltage is increased with the frequency shift, such that the voltage across the TES is constant and therefore the TES power, -resistance and -current are constant.

\begin{figure}[htbp]
	\centering
	\includegraphics[width=\columnwidth]{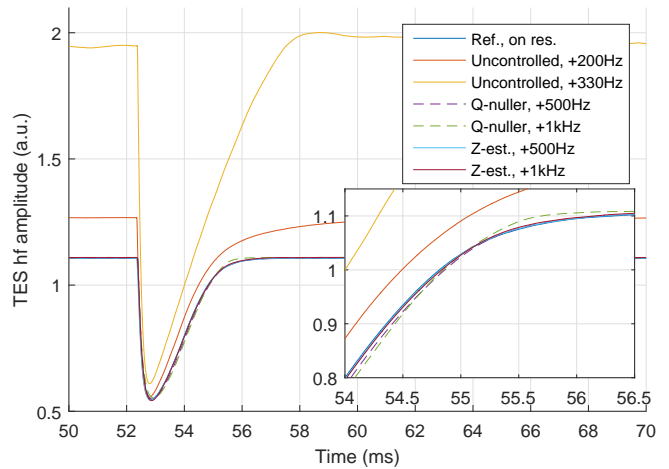}
	\caption{Measured TES current pulse (hf amplitude) in response to an X-ray photon for various frequency shifts and controllers}
	\label{fig:Pulse-response}
\end{figure}

Fig.~\ref{fig:Pulse-response} shows the measured baseband $I$-current (=hf TES current amplitude) in response to an x-ray pulse.
It is clear that a controller is required, because with uncontrolled frequency shift, the response is already much slower for a frequency shift of \unit[200]{Hz}.
The curves for the $Q$-nuller (dashed) are much closer to the reference on-resonance curve.
For a frequency shift of \unit[1]{kHz}, the $Q$-nuller deviates noticeably from the reference in Fig.~\ref{fig:Pulse-response}.
With the Z-estimator, the response is virtually equal to the reference response, making the Z-estimator the better choice.

\begin{figure}[htbp]
	\centering
	\includegraphics[width=\columnwidth]{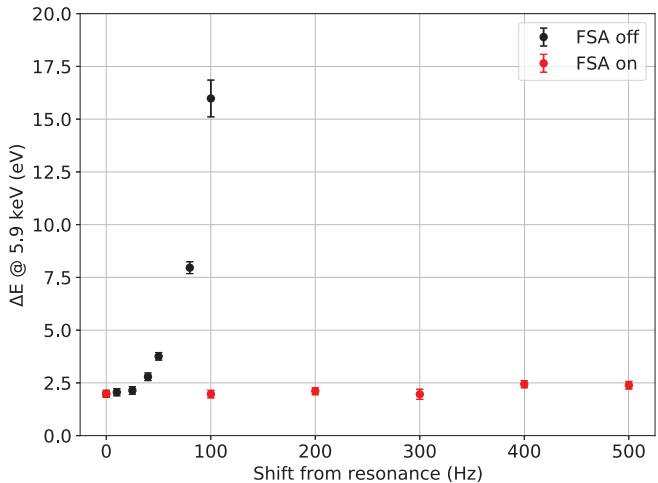}
	\caption{Energy resolution as function of frequency shift with inactive (black) and active (red) controller}
	\label{fig:FSA_freqshift_Paul}
\end{figure}

The slow response of the uncontrolled frequency shifted pixel in figure~\ref{fig:Pulse-response}, translates directly to a reduction of energy resolution, shown in figure~\ref{fig:FSA_freqshift_Paul}.
The purpose of carrier frequency shift is to reduce the energy resolution deterioration as reported by Akamatsu \emph{et al.}\cite{akamatsu_progress_2020}
Figure~\ref{fig:FSA_freqshift_Paul} shows that carrier frequency shift severely compromises the energy resolution when no additional measures are taken (black).
With active $Q$-nuller controller (in red), the energy resolution displays no apparent deterioration.
The controllers thus enable the practical application of frequency shift. 

Extensive results confirming the merits of the new controllers in x-ray spectrometry applications are presented by Vaccaro \emph{et al.}\cite{vaccaro_frequency_2021} for single pixel and in multiplexing situations.
In a three-pixel multiplexing experiment with the $Q$-nuller controller and an LC resonator bandwidth of about \unit[1]{kHz}, a frequency shift of \unit[600]{Hz} did not result in a significant degradation of the energy resolution.
Multipixel stable operation was demonstrated with 22 pixels.
The limitation in the number of pixels was due to the available number of cold components and not algorithm-related.
Multi-pixel operation, preserving the single-pixel \unit[$\approx 2.6$]{eV} resolution of the TESs was demonstrated for up to 20 pixels.

\subsection{Hardware}
The digital electronics is implemented in a Field Programmable Gate Array (FPGA) which has a range of different hardware resources such as look-up tables for logic or memory, flip flops, etc.
This makes comparing the use of resources complicated.
The Z-estimator performs better than the $Q$-nuller, at the cost of more hardware in the FPGA.
Resources for the $Q$-nuller are about \unit[20]{\%} lower than for the Z-estimator.
The tested implementation was not optimized for efficiency.
Some parts are reused by both controllers and additional switching logic is added.
Even with these limitations, this implementation is about half the size of the BBFB filter.

The effect of the controllers on the dynamic range of the ac bias DAC was also investigated. 
It was found that for an average frequency shift of \unit[250]{Hz}, the increase in DAC voltage is only \unit[5]{\%}, due to the orthogonal addition of the bias and controller voltages. 
The controller is therefore not expected to introduce crosstalk at problematic levels.

\section{Conclusions}
A new method has been developed for off-resonance operation of frequency domain multiplexed TES pixels. 
The main characteristic of the method is the use of controllers with the $Q$ current signal from the BBFB filter as error input, and an output voltage which is orthogonal to the TES current.
Because the modulation of the controller output is shared with the bias ac voltage generation, the phase-shift between the carrier and controller signals is not compromised by differences in numerical delay and implementation requires only small modifications of the existing firmware.
Tuning of the controllers is relatively simple because the controllers automatically adapt to the applied frequency shift.
The controller gain settings are not critical in general.
Only the calibration of the offset angle between modulation and demodulation stages requires some attention but is a straightforward operation.

The parasitic off-resonance impedance presented by the other LC resonators in the channel ($Y_\text{par}$) must be compensated to allow proper calibration of the carrier phase.
Right now carrier rotation and $Y_\text{par}$ are fine-tuned using an iterative procedure, but better knowledge of the transfer of the bias and feedback transmission channels should allow us to predict these. 

As reported by Akamatsu \emph{et al.}\cite{akamatsu_progress_2020}, the energy resolution of individual pixels in an FDM readout channel can reduce by more than \unit[2]{eV} due to intermodulation line noise.
The proposed controllers provide a scalable method to enable carrier frequency shift, which restores the single pixel readout performance in FDM multiplexed readout.

\begin{acknowledgments}
This work is part of the research program Athena with project number 184.034.002, which is financed by the Dutch Research Council (NWO).
\end{acknowledgments}

\section*{Data availability}
The corresponding author makes available the data presented in this paper upon reasonable request.

\appendix

\section{Frequency transformation}
The behavior and stability of the controllers in closed loop with the system needs to be evaluated. 
A complication is that the controller works on quasi dc demodulated signals, while the LC resonator operates on hf signals at the carrier frequency, which may be close to, but not equal to the LC resonance frequency.
The frequency transformation in between makes the system non-linear.
It is necessary to remove the frequency transformation to obtain a baseband description of the LC- and BBFB filters.

\subsection{HF model of BBFB filter}
We will first determine a linear hf description of the BBFB filter including the feedback loop.
It is known that the BBFB filter behaves similar to a narrow-band band pass filter.
We will show here that the LC filter and the baseband feedback filter are equivalent.

\begin{figure}[htbp]
	\centering
	\includegraphics[scale=0.8]{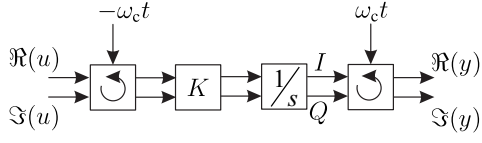}
	\caption{HF forward signal path of the BBFB circuit: an integrator with gain enclosed between clockwise and counterclockwise rotations}
	\label{fig:BBFB_open_loop}
\end{figure}

\begin{figure}[tbp]
	\centering
	\includegraphics[scale=1]{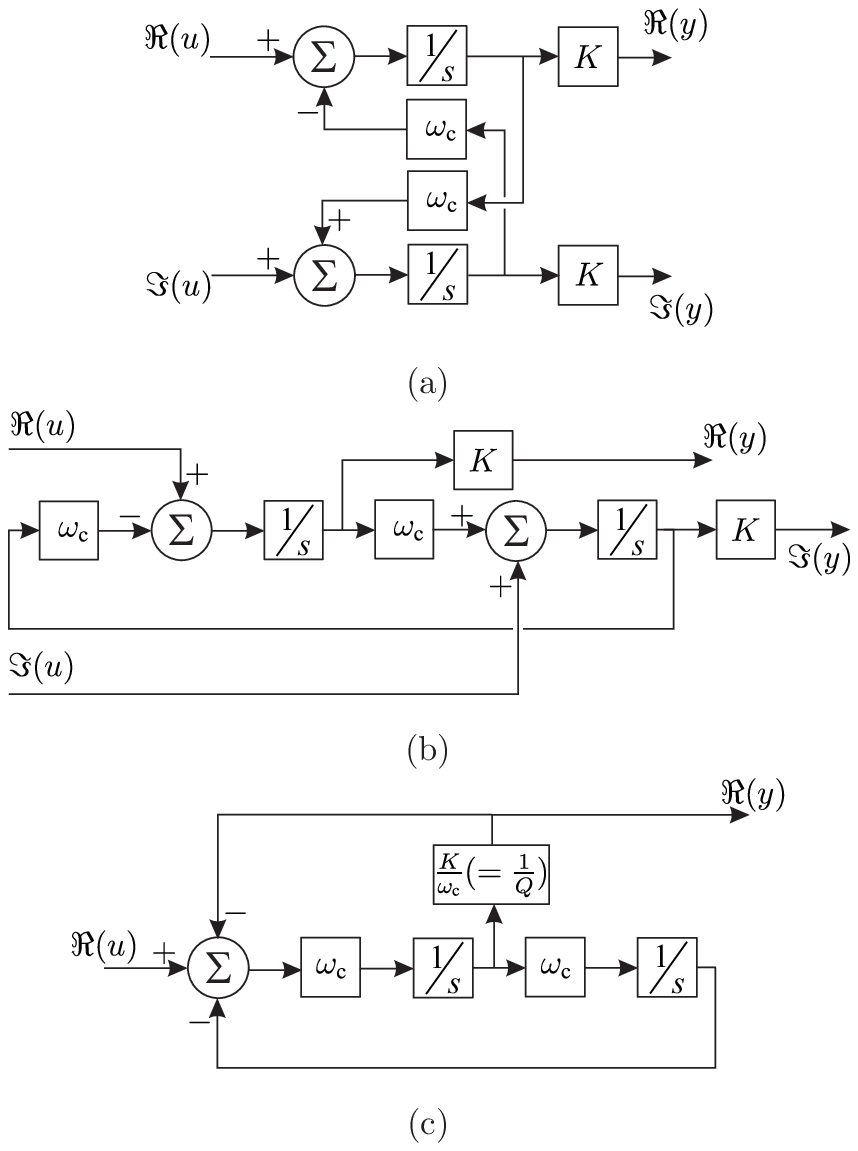}
	\caption{Simplification of the HF model of the BBFB circuit. 
		The shifted integrator of eq.\ \eqref{eq:BBFB_shifted_integrator} is implemented as (a). 
		Reordering exposes a 2\textsuperscript{nd} order loop without damping (b). 
		When adding feedback in the real signal path and ignoring the imaginary in- and output, the band-pass filter with programmable bandwidth (c) results.}
	\label{fig:BBFB_forward}
\end{figure}

The BBFB filter is a hybrid structure with the forward path in baseband- and the feedback path in the hf domain.
The filter, shown in Fig.~\ref{fig:BBFB_open_loop}, consists of an integrator enclosed between two vector rotators, of which typically only the real in- and outputs are used.
When the rotation angle is $\mp \omega_\text{c}t$ respectively, the complex input experiences a frequency transformation of  $\mp \omega_\text{c}$ respectively. 
The frequency shift by the rotators results in an apparent frequency shift of the transfer function in between.
A signal at the input of frequency $+\omega_\text{c}$ experiences a frequency transformation to dc.
This dc signal is integrated to a ramp, and again subjected to a frequency shift of  $+\omega_\text{c}$ back to the original frequency.
The frequency transfer of this filter is the complex function
\begin{align}
	H_\text{BBFB,fw}(s)=\frac{K}{s-j\omega_\text{c}}\ .\label{eq:BBFB_shifted_integrator}
\end{align}
Because this filter has non-real polynomial coefficients in its numerator and/or denominator, it has a single complex pole: the pole is not accompanied by a complex-conjugate pole.
The equivalent linear realization of this filter is shown in Fig.~\ref{fig:BBFB_forward}a.
When the filter is reordered (Fig.~\ref{fig:BBFB_forward}b), it is evident that the system is a 2\textsuperscript{nd} order feedback system.
The baseband feedback loop is closed by feeding back the real output to the real input.
At the same time, the input is multiplied by $\omega_\text{c}$ and the output divided by $\omega_\text{c}$.
The resulting closed loop system in Fig.~\ref{fig:BBFB_forward}c is a 2\textsuperscript{nd} order band-pass filter with Q configurable by forward gain $K$ according to $Q = \frac{\omega_\text{c}}{K}$.
Parameter $K = \frac{\omega_\text{c}}{Q}$ is the resonance bandwidth of the baseband feedback filter.
The transfer function for this circuit for real input signals is 
\begin{align}
	H_\text{BBFB,hf}(s)&=\frac{\frac{K}{\omega_\text{c}^2}s}{\frac{s^2}{\omega_\text{c}^2}+\frac{K}{\omega_\text{c}^2}s+1}\\
	&=\frac{\frac{s}{Q\omega_\text{c}}}{\frac{s^2}{\omega_\text{c}^2}+\frac{s}{Q\omega_\text{c}}+1}
\end{align}

\subsubsection{Equivalence of BBFB and LC filters}
\begin{figure}[tbp]
	\centering
	\includegraphics[scale=0.78]{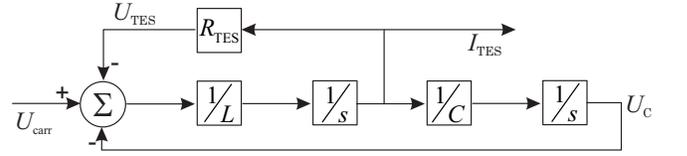}
	\caption{LC filter block diagram}
	\label{fig:LCR-circuit_block_diagram}
\end{figure}

The LC filter circuit of Fig.~\ref{fig:LCR-circuit_landscape} has structure which is very similar to the BBFB filter shown in Fig.~\ref{fig:BBFB_forward}c, as illustrated in Fig.~\ref{fig:LCR-circuit_block_diagram}.
This is confirmed by the transfer function for the LC filter:
\begin{align}
	H_\text{LC,hf}(s)&=\frac{1}{R_\text{TES}}\times\frac{sR_\text{TES}C}{s^2LC+sR_\text{TES}C+1}\\
	&=\frac{1}{R_\text{TES}}\times\frac{\frac{s}{Q_\text{LC}\omega_\text{LC}}}{\frac{s^2}{\omega_\text{LC}^2}+\frac{s}{Q_\text{LC}\omega_\text{LC}}+1}
\end{align}
The only difference is a gain of $\nicefrac{1}{R_\text{TES}}$.

Because the BBFB and LC filter have the same structure, a baseband model with rotators can be created for the LC filter.
Since parameter $K$ is the bandwidth $\frac{\omega}{Q}$, the baseband integrator gain for the LC model can be calculated:
\begin{align}
	K_\text{LC}&=\frac{\omega_\text{LC}}{Q_\text{LC}}\\
	&=\sqrt{\frac{1}{LC}}\times R_\text{TES}\sqrt{\frac{C}{L}}\\
	&=\frac{R_\text{TES}}{L}
\end{align}
The LC filter in Fig.~\ref{fig:LCR-circuit_block_diagram} is thus equivalent to Fig.~\ref{fig:BBFB_open_loop}, with integrator gain $K=\frac{R_\text{TES}}{L}$, and an additional gain of $\frac{1}{R_\text{TES}}$ before the feedback.

\subsection{Complex baseband model of BBFB filter and LC resonator}
\subsubsection{Analytic vs. real signals}
Demodulation and re-modulation of the signal in the BBFB filter is synchronous with the carrier signal.
The frequency-shifted integrator of eq.\ \eqref{eq:BBFB_shifted_integrator} in the BBFB filter amplifies the positive frequency content of the real-valued input signal because its frequency is exactly at the pole of the BBFB filter.
All negative frequency content of the input signal is suppressed by the filter.
The complex baseband input and -output signals are therefore by approximation \emph{analytic} signals, which are complex-valued signals with only positive-frequency content.
The system can thus be analyzed assuming analytic signals.

\begin{figure}[tbp]
	\centering
	\includegraphics[scale=0.8]{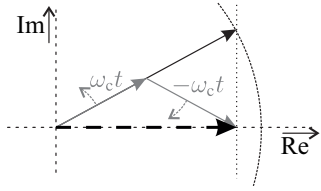}
	\caption{Amplitude halving when reducing a complex signal to its real component. The original analytic signal (black) is reduced to its real part (dashed). This real signal implicitly consists of opposite-frequency components (grey) of half the original amplitude.}
	\label{fig:Complex_to_real_subtractive}
\end{figure}

In the system however, the imaginary component of the complex signal is ignored at the BBFB demodulation input.
The remaining real-valued sinusoidal signal is a sum of complex-conjugate (positive and negative frequency\nobreakdash-) components. Fig.~\ref{fig:Complex_to_real_subtractive} shows that when the real component of a (complex) analytic signal is taken and the imaginary component is ignored, the magnitude of the positive frequency components is halved.
The transfer for analytic signals of the $\Re()$ operator is thus $\nicefrac{1}{2}$.

\begin{figure}[tbp]
	\centering
	\includegraphics[scale=1]{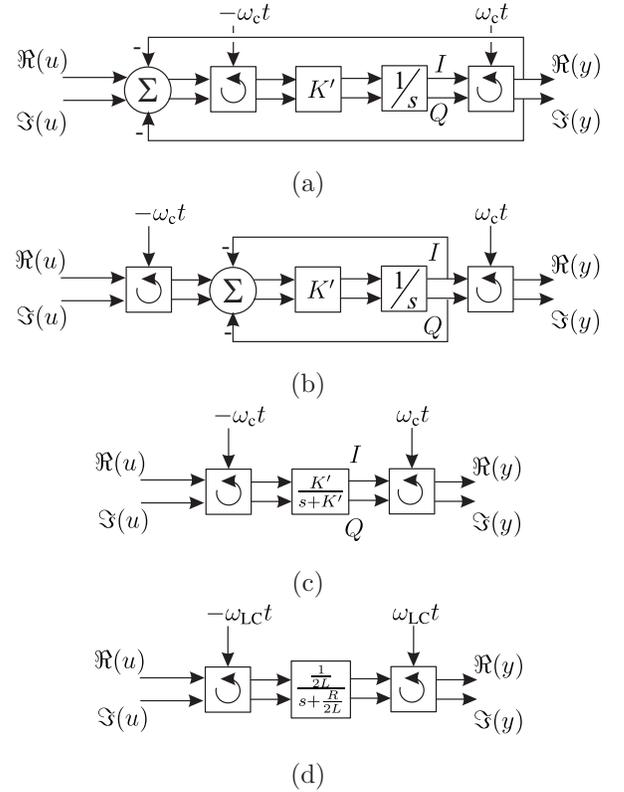}
	\caption{Reduction of the BBFB model with complex feedback ($K'=\nicefrac{K}{2}$). 
		The block model of (a) is the complex equivalent of the implementation in the digital electronics.
		The complex feedback can be moved between the rotation (b).
		The feedback loop between the rotations reduces to a simple low-pass filter for BBFB (c) as well as the LC resonator (d) due to their equivalence.}
	\label{fig:BBFB_complex-dc-feedback}
\end{figure}

In contrast to the actual system, the imaginary output of the LC filter, as well as the imaginary BBFB feedback signals are potentially available in mathematical models of Fig~\ref{fig:BBFB_complex-dc-feedback}.
When the input signal is analytic instead of real, the system gain must be the same to obtain the same response.
The (implicit) complex to real conversion must therefore be replaced by an explicit gain of $\nicefrac{1}{2}$.
The effective BBFB gain $K'$ is therefore: $K'=\nicefrac{K}{2}$ in Fig.~\ref{fig:BBFB_complex-dc-feedback}a through c.
Likewise, the integrator gain in the baseband LC filter model halves from $\frac{R_\text{TES}}{L}$ to $\frac{R_\text{TES}}{2L}$ for analytic signals (Fig.~\ref{fig:BBFB_complex-dc-feedback}d).

With complex feedback, the block diagram of Fig.~\ref{fig:BBFB_complex-dc-feedback}a can be reordered.
The feedback can be moved to between the rotation operations because rotations with $\pm\omega_\text{c}t$ cancel (Fig.~\ref{fig:BBFB_complex-dc-feedback}b).
The feedback around the integrator results in a low pass filter in Fig.~\ref{fig:BBFB_complex-dc-feedback}c.
Analogous, the LC filter is transformed to the subsystem in Fig.~\ref{fig:BBFB_complex-dc-feedback}d.

\subsection{Pixel readout baseband model}
\begin{figure}[tbp]
	\centering
	\includegraphics[scale=1]{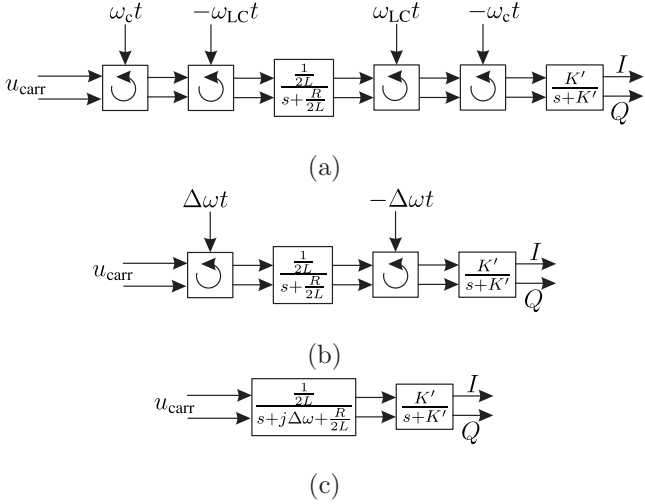}
	\caption{Pixel baseband model assuming analytic signals where $u_\text{carr}$ is the complex input phasor. The original model (a) can be reduced by merging rotations~(b). The remaining rotation is incorporated in the transfer function as a frequency shift~(c).}
	\label{fig:FSA-baseband_reduced-circuits}
\end{figure}

The analytic signal models of the LC filter and the BBFB filter constituting the readout circuit for a single pixel can now be connected to get a model for the complete signal chain from carrier input to measured TES current output (Fig.~\ref{fig:FSA-baseband_reduced-circuits}a).
The input signal is generated by rotating the baseband carrier phasor $u_\text{carr}$ with $\omega_\text{c}t$.
The output modulator in the BBFB filter is discarded because it is no longer used since the feedback was moved between the rotators (Fig.~\ref{fig:BBFB_complex-dc-feedback}b).

The rotations with $\pm\omega_\text{c}t$ and $\mp\omega_\text{LC}t$ before and after the LC filter partly cancel, so the series connection reduces to Fig.~\ref{fig:FSA-baseband_reduced-circuits}b.
The difference between the carrier frequency and the LC resonance frequency is the frequency shift $\Delta\omega=\omega_\text{c}-\omega_\text{LC}$.
The remaining frequency shifts $\pm\Delta\omega$ on either side of the LC filter are incorporated in the LC filter transfer function in Fig.~\ref{fig:FSA-baseband_reduced-circuits}c.

\section{Parallel resonator leakage $Y_\text{par}$ Compensation}
The current measured by the SQUID is the TES current plus the current through all the parallel resonators.
In baseband, the TES current can be reconstructed from the measured (SQUID) current through subtraction of the expected parallel current:
\begin{align}
	\hat{I}_\text{TES}&=\hat{I}_\text{SQUID}-\frac{\vec{U}_\text{ctrl}+U_\text{bias}}{Z_\text{par}}\label{eq:ITES_scientific}
\end{align}
where $Z_\text{par}$ is the cumulative reactive impedance of all the other resonators at the pixel's modulation frequency.
The required control voltage for the Z-estimator controller is:
\begin{align}
	\vec{U}_\text{ctrl}&=Z_\text{LC}\times I_\text{TES}\label{eq:control_voltage}
\end{align}
This constitutes an algebraic loop which we can solve for $\hat{I}_\text{TES}$ by eliminating $\vec{U}_\text{ctrl}$:
\begin{align}
	\hat{I}_\text{TES}&=\hat{I}_\text{SQUID}-\frac{Z_\text{LC}\times \hat{I}_\text{TES}}{Z_\text{par}}-\frac{U_\text{bias}}{Z_\text{par}} \Leftrightarrow\\
	\hat{I}_\text{TES}&=\frac{Z_\text{par}}{Z_\text{par}+Z_\text{LC}}\times\left(\hat{I}_\text{SQUID}-U_\text{bias}\times Y_\text{par}\right)\\
	\hat{I}_\text{TES}&=\frac{Z_\text{par}}{Z_\text{par}+Z_\text{LC}}\times I'_\text{TES}\label{eq:ITES}\\
	\intertext{and thus} \vec{U}_\text{ctrl}&=\frac{Z_\text{par}Z_\text{LC}}{Z_\text{par}+Z_\text{LC}}\times I'_\text{TES}\label{eq:Uctrl}
\end{align}
where $I'_\text{TES}$ is just an intermediate calculation result with no physical meaning.
Using a compensation admittance $Y$ rather than an impedance $Z$ is appropriate to model a parallel circuit such as this.
It also has significant implementation benefits. 
When no compensation is needed, a value of $Y_\text{par}=0$ is applied instead of $Z_\text{par}=\nicefrac{1}{0}$.
Therefore there is little danger of overflow in the compensation value.
Also, the calculation of the compensation value is a multiplication $U_\text{bias}\times Y_\text{par}$ rather than a division $\nicefrac{U_\text{bias}}{Z_\text{par}}$ which would be much more expensive to implement.

The result of $I'_\text{TES}=\hat{I}_\text{SQUID}-U_\text{bias}\times Y_\text{par}$ will be used as input of the controller rather than $\hat{I}_\text{TES}$. 
The multiplier $\frac{Z_\text{par}}{Z_\text{par}+Z_\text{LC}}$ just changes the controller gain by a small amount, which is insignificant since the controller gain value is not critical.
The output of the controller $\hat{Z}$ will be the estimated compensation impedance
\begin{align}
	\hat{Z}=\frac{Z_\text{par}Z_\text{LC}}{Z_\text{par}+Z_\text{LC}}
\end{align}
which allows calculation of the controller voltage with~\eqref{eq:Uctrl}, using the same intermediate result $I'_\text{TES}$.
Interestingly, $\hat{Z}$ is the parallel impedance of \emph{all} the LC resonators at the carrier frequency instead of just the LC series impedance of the pixel's TES sensor, which was the original intention.

The exact value of ${I}_\text{TES}$, which is required for scientific analysis, can now be calculated by filling in the controller voltage calculated with \eqref{eq:Uctrl} in \eqref{eq:ITES_scientific}.		
\begin{align}
	\hat{I}_\text{TES}&=\left(\hat{I}_\text{SQUID}-U_\text{bias}\times Y_\text{par}\right)-\vec{U}_\text{ctrl}\times Y_\text{par}
\end{align}
This way we skip explicit solving of $Z_\text{LC}$.
This last step is required for both the Q-nuller and the Z-estimator.

\section*{References}
%

\end{document}